\documentclass[11pt,a4paper]{article}
\usepackage{color}
\usepackage{jheppub}
\usepackage{fancybox}
\usepackage{amsmath}
\usepackage{amsfonts}
\usepackage{slashed}
\usepackage{amssymb}

\newcommand{\Comment}[1]{{}}

\Comment{\usepackage{color}
\definecolor{MyDarkBlue}{rgb}{0.15,0.15,0.45}
\usepackage[linktocpage=true]{hyperref}
\hypersetup{
colorlinks=true,
citecolor=MyDarkBlue,
linkcolor=MyDarkBlue,
urlcolor=MyDarkBlue,
pdfauthor={Thiago R. Araujo and Horatiu Nastase},
pdftitle={On ${\cal N}=1$ susy backgrounds with AdS factor from nonabelian T-duality},
pdfsubject={hep-th}
}

\usepackage[numbers,sort&compress]{natbib}
\usepackage{hypernat}
}

\newcommand\ignore[1]{}
\def\one{{\,\hbox{1\kern-.8mm l}}}

\def\Tr{{\rm Tr\, }}

\def\a{\alpha}\def\b{\beta}

\newcommand{\Cset}{{\,\,{{{^{_{\pmb{\mid}}}}\kern-.45em{\mathrm C}}}}}

\newcommand{\be}{\begin{equation}}
\newcommand{\bea}{\begin{eqnarray}}

\newcommand{\ee}{\end{equation}}
\newcommand{\eea}{\end{eqnarray}}

\title{\boldmath ON $\mathcal{N}=1$ SUSY BACKGROUNDS WITH $AdS$ FACTOR FROM NONABELIAN T-DUALITY}

\author{THIAGO R. ARAUJO}
\author{and HORATIU NASTASE}

\affiliation{Instituto de F\'{i}sica Te\'{o}rica, UNESP-Universidade Estadual Paulista\\ R. Dr. Bento T. Ferraz 271, Bl. II, Sao Paulo 01140-070, SP, Brazil}

\emailAdd{taraujo@ift.unesp.br}
\emailAdd{nastase@ift.unesp.br}

\abstract{We consider the action of a non-abelian T-duality on backgrounds with an AdS${}_5$ factor in type IIA supergravity, finding a new 
type IIB background, as well as the non-abelian T-dual of a domain wall solution which has as limits the non-abelian T-dual of 
$AdS_5\times T^{1,1}$ and the non-abelian T-dual of $AdS_3\times \mathbb{R}^2\times S^2\times S^3$.
We explore some consequences of non-abelian T-duality for the dual conformal field theories.}


\begin{document}

\maketitle

\section{Introduction}

Since the begining of string theory, the notion of duality symmetries has played an important role. In the early days of string theory, when 
it was a model for strong interactions, the observation that the amplitude of a scattering process could be written equally well in terms of 
the s- or t-channel Mandelstam variables led to the name of "dual models" \cite{Green1987}.

Nowadays, many different dualities exist in string theory, for instance Gauge/Gravity duality \cite{Maldacena1997, Aharony1999a}, 
S-duality, T-duality \cite{Polchinski1998}, Mirror Symmetry \cite{Greene1996, Strominger1996}, Langlands duality \cite{Frenke2005, Frenke2009}. 
In this paper we are interested in the non-abelian generalization of T-duality (started by the paper \cite{de la Ossa:1992vc}), 
which is the case where the isometry group of the background is non-abelian. 
Differently than its abelian cousin, the non-abelian T-duality has been poorly understood, and just recently the action of the transformation on the RR 
fields was found \cite{Sfetsos2010, Itsios2013}. 

The general procedure for T-duality follows the original idea of Buscher \cite{Buscher1988}, that is, we start with a $\sigma$-model which 
supports an isometry such as $U(N)$. Then we gauge the isometry, but we need to impose a constraint by means of Lagrange multipliers 
which guarantees that the connection field strength remains equal to zero. This constraint enforces the condition that after 
gauging the isometry, the initial degrees of freedom remain unchanged.

The duality works as follows. On one hand, by solving the equation of motion for the Lagrange multipliers and replacing the solution into 
the action, we recover the original model. If instead we solve the equation of motion for the connection and we gauge fix, we find the 
dual $\sigma$-model.

Non-abelian T-duality can be used as a solution  generating technique, that is, starting from a solution of supergravity, we can find 
another solution by a simple set of transformations rules. These solutions can be understood better through another type of duality, the
gauge/gravity correspondence.

Particularly interesting solutions in light of this fact are those with $AdS_5$ factors. In fact, one of the first examples of the 
application of a non-abelian T-duality transformation in a background supporting a non-trivial RR field was in the Klebanov-Witten 
solution which consists of a space of the form $AdS_5\times T^{1,1}$, where $T^{1,1}$ is the homogenous space $(SU(2)\times SU(2))/U(1)$.

Recently, \cite{Macpherson2014} reported a large class of new solutions with $AdS_5$ factors and made the analysis of the field theory\footnote{Nonabelian
T-duality on solutions with $AdS$ factors was considered also in \cite{Barranco:2013fza,Lozano:2014ata,Lozano:2012au}.}, 
following \cite{Sfetsos2014}, which performed a non-abelian T-duality in a type $IIB$ solution of the type $AdS_5\times X^5$, 
where this solution was obtained in \cite{Gauntlett2004} after a dimensional reduction of $D=11$ supergravity, followed by an abelian T-duality. 
The study of non-abelian T-duality of AdS backgrounds was initiated in \cite{Itsios2013,Lozano:2013oma}

In this article we explore the non-abelian T-duality on the type IIA supergravity solution (that is, before the abelian T-duality which gives 
$AdS_5\times X^5$) of the form $AdS_5\times_w \mathcal{M}_5$, where the internal manifold is obtained after a dimensional reduction 
of a space that consists of a $2$-sphere bundle over $S^2\times T^2$ \cite{Gauntlett2004}. 
Another application considered relates to the background found in \cite{Gauntlett2014}. It consists of a domain wall with non trivial fluxes 
in the NS-NS and RR sectors. This domain wall solution flows to the background $AdS_3\times \mathbb{R}^2\times S^2\times S^3$ in the IR limit, 
and in the UV to $AdS_5\times T^{1,1}$. We study the T-dual of this domain wall and see that it has as limits the T-dual of $AdS_5\times T^{1,1}$ 
and  $AdS_3\times \mathbb{R}^2\times S^2\times S^3$. We then study the implication of non-abelian T-duality for the dual conformal 
field theories, through a calculation of central charges.

The paper is organized as follows. In section 2 we review non-abelian T-duality and in section 3 we apply it to the warped $AdS_5$ solution.
In section 4 we consider the T-dual of the domain wall solution. In section 5 we consider dual conformal field theory aspects of the T-dual 
solution and calculate central charges, and in section 6 we conclude.

\section{Non-Abelian T-Duality in a nutshell} \label{section-abtd}

Since the present work deals with the uses of the Non-Abelian T-duality as a solution generating technique, we start with a 
review of this procedure, following mostly \cite{Itsios2013}. Consider a background that supports an $SU(2)$-structure, so that we write the metric in the form
\begin{equation}
ds^2=G_{\mu\nu}(x) dx^\mu dx^\nu +2 G_{\mu i}(x) dx^\mu L^i+g_{ij}(x) L^i L^j \label{metric}
\end{equation}
where $\mu,\nu=1,\dots,7$, and $L^i$ are the Maurer-Cartan forms for $SU(2)$. In general we also have non-trivial Kalb-Ramond two-forms 
\begin{equation}
B=B_{\mu\nu} dx^\mu\wedge dx^\nu +B_{\mu i} dx^\mu \wedge L^i+ \frac{1}{2}b_{ij} L^i\wedge L^j, \label{kalbramond}
\end{equation}
and a dilaton $\Phi=\Phi(x)$. The important point here is that all dependence on the $SU(2)$ Euler angles $\theta, \psi, \phi$ is 
contained in the one-forms $L^i$.

Next, define the vielbeins 
\begin{equation}
\begin{split}
\mathfrak{e}^A&= e^A_\mu dx^\mu\\
\mathfrak{e}^a&= \kappa^a_{\phantom{a}j} L^j+\lambda^a_\mu dx^\mu,
\end{split}
\end{equation}
with $A= 1,\dots,7$ and $a=1,2,3$. Imposing
\begin{equation}
 ds^2=\eta_{AB}\mathfrak{e}^A\mathfrak{e}^B+\mathfrak{e}^a\mathfrak{e}^a,
\end{equation} 
by direct comparison with (\ref{metric}) we have
\begin{equation}
G_{\mu\nu}=\eta_{AB} \mathfrak{e}^A\mathfrak{e}^B+K_{\mu\nu},\quad \kappa^a_{\phantom{a}i}
\kappa^a_{\phantom{a}j}= g_{ij}, \quad \kappa^a_{\phantom{a}i}\lambda^a_\mu=G_{\mu i}\;,
\end{equation}
where we defined  $\lambda^a_\mu\lambda^a_\nu=K_{\mu\nu}$.

If we combine the metric and B field into $Q$ and $E$ by
\begin{equation}
\begin{split}
Q_{\mu\nu} &=G_{\mu\nu}+B_{\mu\nu}, \quad Q_{\mu i}=G_{\mu i}+B_{\mu i}\\
Q_{i \mu}&=G_{i \mu}+B_{i \mu},\quad E_{ij}=g_{ij}+b_{ij}\;,
\end{split}
\end{equation}
one can show that the non-abelian T-dual background is
\begin{equation}\label{duality}
\begin{split}
\widehat{Q}_{\mu\nu} &=Q_{\mu\nu}-Q_{\mu i}M_{ij}^{-1}Q_{j\nu}, \quad \widehat{E}_{ij}=M_{ij}^{-1}\\
\widehat{Q}_{\mu i}&=Q_{ \mu j}M_{ji}^{-1}, \qquad \widehat{Q}_{i\mu}=-M_{ij}^{-1}Q_{ j\mu },
\end{split}
\end{equation}
where the matrix $M$ is defined by 
\begin{equation}
M_{ij}=E_{ij}+\a' f_{ij}^{\phantom{ij}k} v_k.
\end{equation}
Here $f_{ij}^{\phantom{ij}k}=\sqrt{2}\epsilon_{ijk}$ are the structure constants of the group $SU(2)$ and $v_i$ are originally 
Lagrange multipliers, now dual coordinates. We can make the scaling $v_i\to\frac{1}{\sqrt{2}}v_i$, so that the dual fields are written as
\begin{equation}
d\hat{s}^2=\widehat{G}_{\mu\nu}(x) dx^\mu dx^\nu +\frac{2}{\sqrt{2}} \widehat{G}_{\mu i}(x) dx^\mu dv^i+\frac{1}{2}\hat{g}_{ij}(x) dv^i dv^j 
\end{equation}
and 
\begin{equation}
\widehat{B}=\widehat{B}_{\mu\nu} dx^\mu\wedge dx^\nu +\frac{1}{\sqrt{2}}\widehat{B}_{\mu i} dx^\mu \wedge dv^i+ \frac{1}{4}\hat{b}_{ij} dv^i\wedge dv^j.
\end{equation}
and dilaton (transformed at the quantum level as usual)
\be
\hat \phi=\phi-\frac{1}{2}\ln\left(\frac{\det M}{\a'^3}\right).
\ee

Besides the spectator fields $x^\mu$, the dual theory depends on $\theta, \psi, \phi, v^i$, so we have too many degrees of freedom. 
We need to impose a gauge fixing in order to remove three of these variables, usually taken to be $\theta=\psi=\phi=0$. Then one finds
\begin{equation}
(M^{-1})^{ij}=\frac{1}{\det M}\left(\det g g^{ij}+y^iy^j-\epsilon^{ijk}g_{kl}y^l\right)
\end{equation}
where we have defined $b_{ij}=\epsilon_{ijk}b_k$ and $y_{i}=b_i+\a' v_i$. For a gauge fixing different than $\theta=\psi=\phi=0$, one defines $\hat v_i
=D_{ji}v^j$, where
\be
D^{ij}=\frac{1}{2}\Tr(\tau^i g \tau^j g^{-1}),\;\;\;
g=e^{\frac{i}{2}\phi\tau_3}e^{\frac{i}{2}\theta\tau_2}e^{\frac{i}{2}\psi\tau_3}\;
\ee
($\tau_i$ are the Pauli matrices) and replaces everywhere $v_i$ by $\hat v_i$.

The dualization acts differently on the left- and the right-movers which produces two different sets of frames 
$\hat{\mathfrak{e}}^i_+$ and $\hat{\mathfrak{e}}^i_-$ that are related by a Lorentz transformation 
$\hat{\mathfrak{e}}^a_+=\Lambda^a_{\phantom{a}b} \hat{\mathfrak{e}}^b_-$. The action on the spinor representation of the Lorentz group 
is given by
\begin{equation}
\Omega^{-1}\Gamma^a\Omega= \Lambda^a_{\phantom{a}b}\Gamma^b.
\end{equation}

Considering the RR sector in the democratic formalism (we consider the fluxes and their Hodges dual as well), we define the polyforms in 
type II supergravity
\begin{equation}
\textbf{IIB:}\ \ P=\frac{e^\phi}{2}\sum_{n=0}^4\slashed{F}_{2n+1}\ ,\quad \textbf{IIA:}\ \ 
\widehat{P}=\frac{e^{\hat{\phi}}}{2}\sum_{n=0}^5\widehat{\slashed{F}}_{2n}
\end{equation}
Then the non-abelian T-dual forms are obtained by the transformation (applied to the non-abelian case by \cite{Sfetsos2014}, following 
work in the abelian case by \cite{Hassan:1999bv})
\begin{equation}
\widehat{P}=P\cdot \Omega^{-1}.\label{omegas}
\end{equation}

\section{Warped $AdS_5$ solution}

Supersymmetric solutions of $D=11$ supergravity of the form $AdS_5\times_w \mathcal{M}_6$, with non-trivial four form flux living in the 
internal Riemann manifold were considered in \cite{Gauntlett2004}. The authors found that the six dimensional Riemannian manifold always 
admits a Killing vector, and that locally, the five-dimensional space orthogonal to the Killing vector is a warped product of a one dimensional 
space  parametrized by the coordinate $y$ and a four-dimensional complex space $\mathcal{M}_4$.

Also, the authors found a large class of regular solutions. One of this solutions, namely $\mathcal{M}_4=S^2\times T^2$ is peculiar. Firstly 
we can reduce on an $S^1$ direction in the torus $T^2$ so that we can obtain a regular solution of type IIA solution of the form, 
$AdS_5\times X^\prime_5$. Moreover, after a T-duality on the other $S^1$ we get a type IIB solution of the form $AdS_5\times X_5$, 
where $X_5$ is a family of Sasaki-Einstein manifolds, and the global aspects of these spaces was studied in \cite{Gauntlett2004a, Gauntlett2004b}.

The type $IIA$ solution of \cite{Gauntlett2004} is of the form
\begin{subequations}
\begin{align}
\frac{1}{R^2}ds^2&=\underbrace{ds^2(AdS_5)+\alpha_1(y) dy^2+\alpha_2(y) dx^2}_{G_{\mu\nu}(x)dx^\mu dx^\nu}
+\underbrace{\beta_1(y)(L^2_1+L^2_2)+\beta_2(y) L_3^2}_{g_{ij}(x)L_iL_j},\label{solution-a}\\
\frac{1}{R^2}B&=\underbrace{\gamma(y) dx\wedge L_3}_{B_{\mu i} dx^\mu\wedge L_i}\\
\phi&=\phi(y)\\
\frac{1}{R^3}F_{4}^{(RR)}&=\eta(y) dy\wedge Vol(S^2)\wedge L_3\label{solution-d}
\end{align}
\end{subequations}
where $L_i=\sigma_i/\sqrt{2}$, with $i=1,2,3$ are the Maurer-Cartan forms of the group $SU(2)$, satisfying
\begin{equation}
dL_i=-\frac{1}{2}\sqrt{2}\epsilon_{ijk}L_j\wedge L_k,
\end{equation} 
with the left invariant forms
\begin{align}
\sigma_1&=\cos\psi d\theta+\sin\psi\sin\theta d\phi\nonumber\\
\sigma_2&=-\sin\psi d\theta+\cos\psi\sin\theta d\phi\\
\sigma_3&=d\psi +\cos\theta d\phi\nonumber.
\end{align}
The coefficients of this solution are given by
\begin{equation}
\begin{split}
\alpha_1(y)&=e^{-6\lambda}\sec^2\zeta,\quad \alpha_2(y)=e^{-6\lambda},\quad \beta_1(y)=\frac{1-cy}{3},\quad \beta_2(y)=\frac{2\cos^2\zeta}{9},\\
\gamma(y)&=-\frac{\sqrt{2}(ca+cy^2-2y)}{6(a-y^2)}\quad\text{and}\quad \eta(y)=-\frac{2\sqrt{2}(1-cy)}{9}=-\frac{2\sqrt{2}}{3}\b_1,
\end{split}
\end{equation}
so that the metric is
\begin{subequations}
\begin{equation}
\begin{split}
ds^2&=R^2ds^2(AdS_5)+R^2e^{-6\lambda}\sec^2\zeta dy^2+R^2e^{-6\lambda} dx^2+R^2\frac{1-cy}{6}(d\theta^2+\sin^2\theta d\phi^2)\\
&+\frac{R^2}{9}\cos^2\zeta (d\psi+\cos\theta d\phi)^2,
\end{split}
\end{equation}
where $x$ parametrizes the circle $S^1$ of length $2\pi\a'/(lR^2)$, with\footnote{At the level of the supergravity action, the periodicity of $x$ is 
arbitrary \cite{Gauntlett2004}.  But it is T-dual to a IIB solution involving Sasaki-Einstein spaces, for which there is a geometric constraint on the 
periodicity \cite{Gauntlett2004a}.} 
\be
l=\frac{q}{3q^2-2p^2+p\sqrt{4p^2-3q^2}}\;,
\ee
$(\theta,\phi)$ are the polar and azimuthal angles in $S^2$, $y\in (y_1,y_2)$ and $0\leq \psi\leq 2\pi$
(note that in our conventions, $x$ and $y$ are dimensionless, i.e. are written in units of $R$). 
The angle $\zeta$ is defined by $\sin \zeta=2y e^{-3\lambda}$ and $e^{6\lambda}=2(a-y^2)/(1-cy)$ and $a,c$ are constants such that, if 
$c\neq 0$ then $0<a<1$, and if $c=0$ then $a\neq 0$, and if $c\neq 0$ one can set it to 1 and find
\be
a=\frac{1}{2}+\frac{3q^2-p^2}{4p^3}\sqrt{4p^2-3q^2}\;,
\ee
where $p,q\in\mathbb{Z}$. 

The dilaton is
\begin{equation}
\phi=-3\lambda
\end{equation}
and the Kalb-Ramond field is
\begin{equation}
B=R^2\frac{(ca+cy^2-2y)}{6(a-y^2)}(d\psi+\cos\theta d\phi)\wedge dx.
\end{equation}
\end{subequations}

In the RR sector, we have only a nonzero four-form field
\begin{equation}
F_{4}=-R^3\frac{2(1-cy)}{9}dy\wedge (d\psi+\cos\theta d\phi)\wedge Vol(S^2).
\end{equation}

In what follows, it is convenient to use the frame fields
\begin{align}
\mathfrak{i}^a&=e^a_{\hat{a}} dx^{\hat{a}}\quad AdS_5\ \text{directions}\nonumber\\
\mathfrak{e}^x&=R\alpha_1^{1/2} dx, \quad \mathfrak{e}^y=R\alpha_2^{1/2} dy\\
\mathfrak{e}^1&=R\beta_1^{1/2} L_1\ ,\quad \mathfrak{e}^2=R\beta_1^{1/2} L_2\ , \quad \mathfrak{e}^3=R\beta_2^{1/2} L_3,\nonumber
\end{align}
so that we have the matrix ${\kappa^a}_j$ given by
\begin{equation}
\kappa=\begin{pmatrix}
R\beta_1^{1/2} & 0 & 0 \\ 
0 & R\beta_1^{1/2} & 0 \\ 
0 & 0 & R\beta_2^{1/2}
\end{pmatrix} .
\end{equation}

\subsection{Nonabelian T-dual model}

We want to T-dualize \cite{Itsios2013} (see also \cite{Gevorgyan2013} for the complete list of dual transformations) with respect to the 
$SU(2)$. As in section \ref{section-abtd}, we form the matrix $M_{ij}$, given by $M_{ij}=g_{ij}+b_{ij}+\a'\epsilon_{ijk}\hat v_k$, so 
($b_{ij}=0$, $g_{ij}={\kappa^a}_i{\kappa^a}_j$),
\begin{subequations}
\begin{equation}
M=\begin{pmatrix}
R^2\beta_1 & \a\hat{v}_3 & -\a'\hat{v}_2 \\ 
-\a'\hat{v}_3 & R^2\beta_1 & \a'\hat{v}_1 \\ 
\a'\hat{v}_2 & -\a'\hat{v}_1 & R^2\beta_2
\end{pmatrix}.
\end{equation} 
We pick a gauge where $\theta=\phi=v_2=0$, so that $\hat{v}=(\cos \psi v_1, \sin\psi v_1, v_3)$. This gauge is useful when the vector 
$\partial_\psi$ is a Killing vector as the present case (see \cite{Itsios2013}, for further possible choices). Therefore, the matrix $M$ in this gauge is
\begin{equation}
M=\begin{pmatrix}
R^2\beta_1 & \a'v_3 & -\a'\sin\psi v_1 \\ 
-\a' v_3 & R^2\beta_1 & \a' \cos\psi v_1 \\ 
\a'\sin\psi v_1 & -\a' \cos\psi v_1 & R^2\beta_2
\end{pmatrix}.
\end{equation}
\end{subequations}
The dilaton in the dual theory is given by
\begin{equation}
\widehat{\phi}=\phi-\frac{1}{2}\ln \left(\frac{\Delta}{\a'^3}\right),
\end{equation}
where $\Delta\equiv\det M=R^2[(R^4\beta_1^2+\a'^2v_3^2)\beta_2+\a'^2v_1^2\beta_1]$. 

To simplify the notation, from now on we absorb $R^2$ in $\b_1,\b_2$, 
$\a'$ in $v_1,v_3$, as well as $R^2$ in $\a_1,\a_2,\gamma$.

The inverse of the matrix $M$ is then
\begin{equation}
(M^{-1})^T=\frac{1}{\Delta}\begin{pmatrix}
\beta_1\beta_2+v_1^2\cos^2\psi\quad &v_3\beta_2+v_1^2\cos\psi\sin\psi\quad & v_1v_3\cos\psi-v_1\beta_1\sin\psi \\ 
-v_3\beta_2+v_1^2\cos\psi\sin\psi & \beta_1\beta_2+v_1^2\sin^2\psi & v_1\beta_1\cos\psi+v_1v_3\sin\psi \\ 
v_1v_3\cos\psi+v_1\beta_1\sin\psi & -v_1\beta_1\cos\psi+v_1v_3\sin\psi & v_3^2+\beta_1^2
\end{pmatrix}.\label{invmatrix}
\end{equation}

Finally, taking the symmetric and skew-symmetric part of (\ref{duality}), we get the following T-dual fields
\begin{center}
\shadowbox{\begin{minipage}{3.7in}
$$\widehat{G}_{\mu\nu}=G_{\mu\nu}-\frac{1}{2}\left(Q_{\mu i}M^{-1}_{ij}Q_{j \nu}+Q_{\nu i}M^{-1}_{ij}Q_{j \mu}\right) $$
$$\widehat{G}_{\mu i}=\frac{1}{2}\left(Q_{\mu j}M^{-1}_{ji}-Q_{j \mu}M^{-1}_{ij}\right) $$
$$\hat{g}_{ij}=\frac{1}{2}\left(M^{-1}_{ij}+M^{-1}_{ji}\right)$$
$$\widehat{B}_{\mu\nu}=B_{\mu\nu}-\frac{1}{2}\left(Q_{\mu i}M^{-1}_{ij}Q_{j \nu}-Q_{\nu i}M^{-1}_{ij}Q_{j \mu}\right) $$
$$\widehat{B}_{\mu i}=\frac{1}{2}\left(Q_{\mu j}M^{-1}_{ji}+Q_{j \mu}M^{-1}_{ij}\right) $$
$$\hat{b}_{ij}=\frac{1}{2}\left(M^{-1}_{ij}-M^{-1}_{ji}\right)$$
\end{minipage}}
\end{center}

For the solution (\ref{solution-a} - \ref{solution-d}), where $x^\mu=\{x,y,AdS_5$ coordinates$\}$ and $i=1,2,3$, we consider just the terms 
which will be affected by the non-abelian T-duality, namely, $Q_{x x}, Q_{x i}$ and $Q_{ij}$, giving
$$
\begin{array}{|l|l|}
  \hline
  Q_{x x}=G_{x x}=\alpha_2(y) & Q_{x 3}=B_{x 3}=\gamma(y)\\
  \hline \hline
  Q_{11}=Q_{22}=g_{11}=\beta_1(y) &   Q_{33}=g_{33}=\beta_2(y) \\
  \hline
 \end{array}
$$
For the metric, we obtain $\widehat{G}_{\mu\nu}=G_{\mu\nu},\, \widehat{G}_{\mu i}=0\,  \forall\, \mu, \nu\neq x$. Moreover, we have the 
diagonal component
\begin{equation}
\widehat{G}_{xx}=\alpha_2(y)+\frac{1}{\Delta}(v_3^2+\beta_1^2)\gamma^2,
\end{equation}
the crossed terms
\begin{align}
\widehat{G}_{x1}&=\frac{1}{\Delta}\gamma v_1v_3\cos\psi\nonumber\\
\widehat{G}_{x2}&=\frac{1}{\Delta}\gamma v_1v_3\sin\psi\\
\widehat{G}_{x3}&=\frac{1}{\Delta}\gamma (v_3^2+\beta_1^2)\;,\nonumber
\end{align}
and the $g_{ij}$ components
\begin{align}
\hat{g}_{11}=\frac{1}{\Delta}(\beta_1\beta_2+v_1^2\cos^2\psi),\quad \hat{g}_{12}&
=\frac{1}{\Delta}v_1^2\cos\psi\sin\psi,\quad \hat{g}_{13}=\frac{1}{\Delta}v_1v_3\cos\psi \nonumber\\
\hat{g}_{21}=\frac{1}{\Delta}v_1^2\cos\psi\sin\psi ,\quad \hat{g}_{22}&
=\frac{1}{\Delta}(\beta_1\beta_2+v_1^2\sin^2\psi) ,\quad \hat{g}_{23}=\frac{1}{\Delta}v_1v_3\sin\psi \\
\hat{g}_{31}=\frac{1}{\Delta}v_1 v_3\cos\psi ,\quad \hat{g}_{32}&
=\frac{1}{\Delta}v_1 v_3\sin\psi ,\quad \hat{g}_{33}=\frac{1}{\Delta}(v_3^2+\beta_1^2). \nonumber
\end{align}
All in all, we have the type IIB metric 
\begin{equation}
d\hat{s}^2=d\tilde{s}^2+\frac{1}{\Delta}d\Sigma^2,
\end{equation}
where
\begin{subequations}
\begin{equation}
d\tilde{s}^2=ds^2_{AdS}+\alpha_1(y)dy^2+\alpha_2(y)dx^2
\end{equation}
and
\begin{align}
&d\Sigma^2=\gamma^2(v_3^2+\beta_1^2) dx^2+\frac{2 \gamma}{\sqrt{2}}dx
\left[ v_1 v_3(\cos\psi d\hat{v}_1+\sin\psi d\hat{v}_2)+(v_3^2+\beta_1^2)d\hat{v}_3\right]\nonumber\\
&+\frac{1}{2}\left[(\beta_1\beta_2+v_1^2\cos^2\psi)d\hat{v}_1^2
+(\beta_1\beta_2+v_1^2\sin^2\psi)d\hat{v}_2^2+2v_1^2\cos\psi\sin\psi d\hat{v}_1 d\hat{v}_2\phantom{\frac{1}{1}}\right.\\
&\left.\phantom{\frac{1}{1}}+2v_1v_3\cos\psi d\hat{v}_1 d\hat{v}_3+2v_1v_3\sin\psi d\hat{v}_2 d\hat{v}_3+(v_3^2+\beta_1^2)d\hat{v}_3^2\right].\nonumber
\end{align}
Remembering that $\hat{v}=(v_1\cos\psi,v_1\sin\psi,v_3)$, we rewrite it as
\begin{equation}
\begin{split}
d\Sigma^2=&\gamma^2(v_3^2+\beta_1^2) dx^2+\frac{2\gamma}{\sqrt{2}}dx
\left( v_1 v_3dv_1+(v_3^2+\beta_1^2)dv_3\right)+\frac{1}{2}\beta_1\beta_2 v_1^2 d\psi^2+\\
&+\frac{1}{2}(\beta_1\beta_2+ v_1^2)dv_1^2+v_1v_3 dv_1 dv_3+\frac{1}{2}(v_3^2+\beta_1^2)dv_3^2.
\end{split}
\end{equation}
\end{subequations}

For later use, we calculate $\sqrt{\det g_{int}}$ for this metric, where $g_{int}$ refers to the internal, non-AdS, part of the metric. Writing explicitly the 
factors of $R$ and $\a'$, we obtain
\be
\sqrt{g_{int}}=\frac{1}{\Delta^2}R^3\a'^3\sqrt{\a_1}\sqrt{\b_1\b_2}\frac{v_1}{\sqrt{2}}\sqrt{\det \tilde M}\;,
\ee
where $\tilde M$ is the matrix
\be
\tilde M=\begin{pmatrix} \Delta R^2\a_2+\gamma^2 R^4(\a'^2 v_3^2+\b_1^2 R^4)& \frac{\gamma}{\sqrt{2}}R^2\a'^2 v_1v_3& \frac{\gamma}{\sqrt{2}}
R^2(\a'^2  v_3^2+\b_1^2R^4)\\
 \frac{\gamma}{\sqrt{2}}R^2\a'^2 v_1v_3& \frac{\b_1\b_2 R^4+\a'^2 v_1^2}{2}& \a'^2\frac{v_1v_3}{2}\\
\frac{\gamma}{\sqrt{2}}R^2(\a'^2  v_3^2+\b_1^2R^4)&  \a'^2\frac{v_1v_3}{2}& \frac{\a'^2 v_3^2+\b_1^2 R^4}{2}\end{pmatrix}\;
\ee
and we find
\be
\det \tilde M=\frac{\a_2\b_1 R^4}{4}\Delta^2\Rightarrow \sqrt{\det g_{int}}=\frac{R^5\a'^3}{\Delta}\sqrt{\a_1\a_2}\b_1\sqrt{\b_2}\frac{v_1}
{2\sqrt{2}}.\label{detads5}
\ee

Finally, the T-dual Kalb-Ramond field is given by
\bea
\widehat{B}&=&\frac{\gamma v_1\beta_1}{\sqrt{2}\Delta} dx\wedge(-\sin\psi d\hat{v}_1+\cos\psi d\hat{v}_2)\cr
&&+\frac{1}{2\Delta}(-v_3\beta_2d\hat{v}_1\wedge d\hat{v}_2+v_1\beta_1\sin\psi 
d\hat{v}_1\wedge d\hat{v}_3-v_1\beta_1\cos\psi d\hat{v}_2\wedge d\hat{v}_3)\cr
&=&\frac{1}{\Delta}\left[\frac{ v_1^2\beta_1}{\sqrt{2}}\left( \gamma dx+\frac{1}{\sqrt{2}}  
dv_3\right)-\frac{1}{2}v_1 v_3\beta_2 dv_1\right]\wedge d\psi.\label{kr-field}
\eea
The T-dual vielbeins are \footnote{In fact, we have two different sets of dual frame fields related by a Lorentz transformation, 
that is, $\hat{\mathfrak{e}}_+=\Lambda \hat{\mathfrak{e}}_-$, as a result of the different transformation rules of the left- and the 
right- movers in the sigma model \cite{Itsios2013}. For simplicity, in this letter we consider just the $\hat{\mathfrak{e}}_+$ terms.}
\begin{subequations}
\begin{align}
\hat{\mathfrak{e}}^{\prime}_1&=-\frac{\sqrt{\beta_1}}{\sqrt{2}\Delta}\left( v_1 v_3 \beta_2 d\psi
+(v_1^2+\beta_1\beta_2)dv_1+v_1v_3 dv_3\right)-\frac{\gamma \sqrt{\beta_1}}{\Delta}v_1v_3dx\\
\hat{\mathfrak{e}}^{\prime}_2&=-\frac{\sqrt{\beta_1}}{\sqrt{2}\Delta}\left( v_1\beta_1 \beta_2 d\psi-\beta_2 
v_3 dv_1+v_1\beta_1 dv_3\right)-\frac{\gamma \sqrt{\beta_1}}{\Delta}v_1\beta_1dx\\
\hat{\mathfrak{e}}_3 &=-\frac{\sqrt{\beta_2}}{\sqrt{2}\Delta} \left(-v_1^2\beta_1 d\psi+v_1v_3 dv_1
+(v_3^2+\beta_1^2)dv_3\right)-\frac{\gamma\sqrt{\beta_2}}{\Delta}(v_3^2+\beta_1^2)  dx,
\end{align}
where we have defined the rotated vielbeins
\begin{equation}
\begin{pmatrix}
\hat{\mathfrak{e}}^{\prime}_1 \\ 
\hat{\mathfrak{e}}^{\prime}_2
\end{pmatrix} = 
\begin{pmatrix}
\cos\psi & \sin\psi \\ 
-\sin\psi & \cos\psi
\end{pmatrix} 
\begin{pmatrix}
\hat{\mathfrak{e}}_1 \\ 
\hat{\mathfrak{e}}_2
\end{pmatrix}.
\end{equation}
\end{subequations}

In term of this basis we write the Kalb-Ramond field (\ref{kr-field}) as
\begin{equation}
-\frac{\widehat{B}}{2}=-\frac{v_3}{\beta_1}\hat{\mathfrak{e}}_1^{\prime}\wedge \hat{\mathfrak{e}}_2^{\prime}+
\frac{v_1}{(\beta_1\beta_2)^{1/2}}\hat{\mathfrak{e}}_3\wedge \hat{\mathfrak{e}}_2^{\prime}.
\end{equation}

Using these results,we are able to find the RR forms in this type IIB background. We write the four-form 
(\ref{solution-d}) as ($\mathfrak{e}_1\wedge \mathfrak{e}_2\wedge \mathfrak{e}_3=\frac{\b_1\sqrt{\b_2}}{2}vol(S^2)\wedge L_3$, 
remembering that $\b_i$ contain $R^2$)
\begin{equation}
F_{4}=\Xi_0\ dy\wedge \mathfrak{e}_1\wedge \mathfrak{e}_2\wedge \mathfrak{e}_3\equiv G_1^{(3)} 
\wedge \mathfrak{e}_1\wedge \mathfrak{e}_2\wedge \mathfrak{e}_3\;,
\end{equation}
where $G_1^{(3)}=\Xi_0 dy$ with $\Xi_0=-4\sqrt{2}R/(3 \beta_2^{1/2})=4\sqrt{2}/\sqrt{3(1-cy)}$. 
In this way we have written the RR 4-form in the way suited to 
apply the nonabelian T-duality as described in the Appendix.

Using these rules, we find $\hat F_4=\hat F_2=0$ and (reintroducing all factors of $R$ and $\a'$)
\bea
\widehat{F}_1&=&-e^{\phi-\hat \phi}A_0G_1^{(3)}=d \widehat{C}_0=\frac{R^3}{\a'^{3/2}}\frac{4\sqrt{2}}{3}\beta_1 dy\\
\widehat{F}_3&=& d \widehat{C}_2-\widehat{C}_0 d\widehat{B}
=\frac{1}{2}e^{\phi-\hat \phi}G_1^{(3)}\wedge \epsilon^{abc}A_c\hat{\mathfrak{e}}_a\wedge \hat{\mathfrak{e}}_b\nonumber\\
&=&\Xi_0\frac{1}{2}\epsilon^{abc}\mathcal{A}^a dy\wedge \hat{\mathfrak{e}}_b\wedge \hat{\mathfrak{e}}_c\nonumber\\
&=&R^5\sqrt{\a'}\frac{4\sqrt{2}}{3\Delta}\beta_1 dy \wedge\left[\frac{v_1^2\beta_1}{\sqrt{2}}\left(\frac{1}{\sqrt{2}}dv_3
+R^2\gamma dx \right)-\frac{v_1 v_3\beta_2}{2} dv_1\right]\wedge d\psi\nonumber\\
&=&-\frac{1}{\a^{3/2}}\frac{4\sqrt{2}}{3}\frac{1}{\beta_2^{1/2}}dy\wedge \left(\beta_2^{1/2}v_3 \mathfrak{e}_1^{\prime}
\wedge \mathfrak{e}_2^{\prime}+\beta_1^{1/2}v_1 \mathfrak{e}_2^{\prime}\wedge \mathfrak{e}_3\right)\nonumber\\
&=&\hat B\wedge \hat F_1\;,
\eea
where the coefficients from the appendix are
\begin{equation}
A_a=\frac{1}{\Delta^{1/2}}\mathcal{A}_a,
\end{equation}
and $\mathcal{A}_a={\kappa^a}_i\hat v^i=R\a'
(\beta_1^{1/2}v_1\cos\psi, \beta_1^{1/2}v_1\sin\psi, \beta_2^{1/2}v_3)$. This background is supplemented by the 
forms $\widehat{F}_9=\star \widehat{F}_1$ and $\widehat{F}_7=-\star \widehat{F}_3$. Using these expressions it is straightforward to verify that 
the Bianchi identities $dF_1=0$ and $dF_3=H\wedge F_1$ are satisfied. Moreover,  $B\wedge F_3=0$.

For later use, we also compute the Page charges in this geometry. The quantized Page charges in this background are given by 
\footnote{Note that $2\kappa_{10}^2=(2\pi)^7\a'^4$ and $T_{Dp}=(2\pi)^{-9}\a'^{-\frac{p+1}{2}}$, so $2\kappa_{10}^2T_{Dp}=(2\pi l_s)^{7-p}$.}
\bea
\mathcal{Q}^{Page}_{D3}&=&\frac{1}{2\kappa_{10}^2T_{D3}}\int_{\Sigma_5}(\hat F_5-\hat B\wedge \hat F_3)=0\cr
\mathcal{Q}^{Page}_{D5}&=&\frac{1}{2\kappa_{10}^2T_{D5}}\int_{\Sigma_3}(\hat F_3-\hat B\wedge \hat F_1)=0\cr
\mathcal{Q}^{Page}_{D7}&=&\frac{1}{2\kappa_{10}^2T_{D7}}
\int_{y_1}^{y_2}\widehat{F}_1=\frac{R^3}{\a'^{3/2}}\frac{4\sqrt{2}}{9}(y_2-y_1)\left(1-\frac{c(y_1+y_2)}{2}\right)=N_{D7}\;
\eea
where, since after an abelian T-duality along the $x$-direction on the solution (\ref{solution-a}-\ref{solution-d}) 
we get a the Sasaki-Einstein manifold, we have  \cite{Gauntlett:2006ai,Gauntlett2004a}
\bea
y_1&=&\frac{1}{4p}(2p-3q-\sqrt{4p^2-3q^2})\cr
y_2&=&\frac{1}{4p}(2p+3q-\sqrt{4p^2-3q^2})\;,
\eea
the solutions to $\cos^2\zeta=0$,
and $p,q\in \mathbb{N}$ with $(p,q)=1$ for $p>q$.
One may verify that this new background has $\mathcal{N}=1$ supersymmetry, under the criteria of \cite{Itsios2013}. In fact, in \cite{Kelekci2014} 
the authors have proved that the vanishing of the Kosmann derivative in the dualized directions of the Killing spinors means supersymmetry
is preserved.\footnote{The supersymmetry preservation under nonabelian T-duality was discussed before in 
\cite{Barranco:2013fza}  and \cite{Macpherson:2013zba}.} 
In the present case, the derivative trivially vanishes, because the Killing spinors are independent of the dualized directions.
Moreover, in \cite{Kelekci2014} a proof was given for the formula (\ref{omegas}), with closed expressions for the dual $p$-form potentials, that 
can be applied more easily to specific cases.
\vspace{0.5cm}

Note that we could have considered the same calculation with  a different gauge fixing for the Lagrange multipliers. 
Consider that the matrix $M$ is instead
\begin{equation}
M=\begin{pmatrix}
\beta_1 & v_3 & -v_2 \\ 
-v_3 & \beta_1 & v_1 \\ 
v_2 & -v_1 & \beta_2
\end{pmatrix},
\end{equation} 
with $v=(\rho \cos\zeta \sin\chi, \rho \cos\zeta \sin\chi, \rho \cos\chi)$. In this coordinate system, we have that 
$\Delta=\beta_2(\beta_1^2+\rho^2\cos^2\chi)+\beta_1^2\rho^2 \sin^2\chi$. The inverse of the matrix M gives equation (\ref{invmatrix}), 
but with the replacements 
\begin{equation}
\psi\leadsto \zeta,\quad v_1\leadsto \rho\sin\chi,\quad v_2\leadsto \rho\cos\chi.
\end{equation}

\section{Flowing from $AdS_5$ to $AdS_3$}

In a recent paper \cite{Gauntlett2014}, the authors considered the construction of a supersymmetric domain wall that approaches 
$AdS_5\times T^{1,1}$ in the UV limit, and $AdS_3\times \mathbb{R}^2\times S^2\times S^3$ in the IR limit. In this section we consider
the non-abelian T-dual solution of the domain wall ansatz and see that it has as its limit the non-abelian T-dual of the $AdS_5\times T^{1,1}$ and 
$AdS_3\times \mathbb{R}^2\times S^2\times S^3$ in the UV and IR respectively.

In fact, the non-abelian T-dual solution of $AdS_5\times T^{1,1}$ is already known from \cite{Itsios2013}. We therefore start with a short review 
of this solution. We consider the conventions of \cite{Gauntlett2014}. Then the type IIB solution is
\begin{subequations}
\begin{align}
\frac{1}{R^2}ds^2_{AdS_5\times T^{1,1}}&=ds^2_{AdS_5}+\frac{1}{6}(ds_1^2+ds_2^2)+\frac{1}{9}(d\psi+P)^2\\
\frac{1}{R^4}F_5&=4 (vol_{AdS_5}+vol_{T^{1,1}}),
\end{align}
\end{subequations}
and $B=0$, $\phi=$constant, 
where $ds_i^2=d\theta_i^2+\sin^2\theta_i d\phi_i^2$ and $P=\cos\theta_1 d\phi_1+\cos\theta_2 d\phi_2$ and we make the replacements 
$v_1\leadsto 2 y_1$ and $v_3\leadsto 2 y_2$. 
The NS-NS sector of the T-dual background is given by 
\begin{subequations}
\begin{align}
d\hat{s}^2_{T(AdS_5\times T^{1,1})}&=ds^2_{AdS_5}+\lambda_0^2 ds_1^2+\frac{\lambda_0^2\lambda^2}{\Delta}y_1^2\sigma_{\hat{3}}^2\nonumber\\
&\quad+\frac{1}{\Delta}\left[(y_1^2+\lambda^2\lambda_0^2)dy_1^2+(y_2^2+\lambda_0^4)dy_2^2+2y_1 y_2 dy_1 dy_2\right]\\
\widehat{B}&=-\frac{\lambda^2}{\Delta}\left[y_1 y_2dy_1+(y_2^2+\lambda_0^4)dy_2\right]\wedge \sigma_{\hat{3}},\\
e^{-2\hat{\phi}}&=8\Delta\a'^{-3/2},
\end{align}
\end{subequations}
where $\lambda_0^2=1/6$, $\lambda^2=1/9$, $\sigma_{\hat{3}}=d\psi+\cos\theta_1 d\phi_1$, and 
\bea
\hat\Delta&\equiv&\det M=8\Delta=8[\lambda_0^2 y_1^2+\lambda^2(y_2^2+\lambda_0^4)]\cr
&=&\beta_1 v_1^2+\beta_2(v_3^2+\beta_1^2).
\eea
Here $\beta_1=2\lambda_0^2$, $\beta_2=2\lambda^2$, $v_1=2y_1$ and $v_3=2y_2$, and as in section 2, we have absorbed a factor of $R^2$
in $\b_1,\b_2$, and a factor of $\a'$ in $v_1,v_3$. The RR-sector is given by
\begin{equation}
\begin{split}
\a'^{3/2}R\widehat{F}_2&=8\sqrt{2}\lambda_0^4\lambda \sin \theta_1 d\phi_1\wedge d\theta_1 \\
\a'^{3/2}R\widehat{F}_4&=-8\sqrt{2}\lambda_0^4\lambda \frac{y_1}{\Delta}\sin \theta_1 d\phi_1\wedge d\theta_1\wedge \sigma_{\hat{3}}\wedge(\lambda_0^2 y_1 dy_2-\lambda^2 y_2 dy_1).
\end{split}
\end{equation}
For completeness, the T-dual vielbeins are given by
\begin{subequations}
\begin{align}
\hat{\mathfrak{e}}^{\prime}_1&=-\frac{\lambda_0}{\Delta}\left[ (y_1^2+\lambda^2\lambda_0^2)dy_1+y_1 y_2(dy_2+\lambda^2 \sigma_{\hat{3}})\right]\\
\hat{\mathfrak{e}}^{\prime}_2&=\frac{\lambda_0}{\Delta}\left[ \lambda^2y_2dy_1-\lambda^2_0 y_1(dy_2+\lambda^2 \sigma_{\hat{3}})\right]\\
\hat{\mathfrak{e}}_3 &=-\frac{\lambda}{\Delta}\left[ y_1 y_2dy_1+(y_2^2+\lambda_0^4)dy_2-\lambda_0^2 y_1^2\sigma_{\hat{3}}\right],
\end{align}
\end{subequations}
and as before, we defined the rotated vielbeins
\begin{equation}
\begin{pmatrix}
\hat{\mathfrak{e}}^{\prime}_1 \\ 
\hat{\mathfrak{e}}^{\prime}_2
\end{pmatrix} = 
\begin{pmatrix}
\cos\psi & \sin\psi \\ 
-\sin\psi & \cos\psi
\end{pmatrix} 
\begin{pmatrix}
\hat{\mathfrak{e}}_1 \\ 
\hat{\mathfrak{e}}_2
\end{pmatrix}.
\end{equation}

This completes the type IIA background T-dual to $AdS_5\times T^{(1,1)}$ in type IIB supergravity.

\subsection{$AdS_3$ solution and its non-abelian T-dual}

The solution with metric $AdS_3\times \mathbb{R}^2\times S^2\times S^3$ is given by
\begin{subequations}
\begin{align}
\frac{1}{R^2}ds^2_{AdS_3\times \mathbb{R}^2\times S^2\times S^3}
=&\frac{1}{3\sqrt{3}}\left(2ds^2_{AdS_3}+dz_1^2+dz_2^2+ds_1^2+ds_2^2+\frac{1}{2}(d\psi+P)^2\right)\\
\frac{1}{R^2}B=&\frac{-\tau}{6\sqrt{6}}z_1 (vol_1-vol_2)\equiv \frac{-\tau\tilde\b_2}{2\sqrt{2}R^2}z_1 (vol_1-vol_2)\\
\frac{1}{R^2}F_3=&\frac{\tau}{6\sqrt{6}}dz_2 \wedge (vol_1-vol_2)\\
\frac{1}{R^4}F_5=&\frac{1}{27}\left\{ vol_{AdS_3}\wedge \left[4 dz_1\wedge dz_2
+\frac{\tau^2}{2}(vol_1+vol_2)\right]\phantom{\frac{1}{1}}\right.\nonumber\\
&\left.+(d\psi+P)\wedge \left[  vol_1\wedge vol_2+\frac{\tau^2}{8} dz_1\wedge dz_2\wedge (vol_1+vol_2)\right]\right\},
\end{align}
\end{subequations}
where $\tau$ is a constant. 

In order to find its T-dual, we consider the Maurer-Cartan forms
\begin{align}
L_1&=\frac{1}{\sqrt{2}}(\cos\psi d\theta_2+\sin\psi\sin\theta_2 d\phi_2)\nonumber\\
L_2&=\frac{1}{\sqrt{2}}(-\sin\psi d\theta_2+\cos\psi\sin\theta_2 d\phi_2)\\
L_3&=\frac{1}{\sqrt{2}}(d\psi +\cos\theta_2 d\phi_2)\nonumber,
\end{align}
such that $vol_2=2 L_1\wedge L_2$. Using the set-up of section \ref{section-abtd},  the vielbeins related to the directions to be T-dualized are 
\begin{subequations}
\begin{align}
\mathfrak{e}^1&=\tilde{\beta}_1^{1/2} L_1\\
\mathfrak{e}^2&=\tilde{\beta}_1^{1/2} L_2\\
\mathfrak{e}^3&=\tilde{\beta}_2^{1/2} (L_3+1/\sqrt{2}\cos\theta_1 d\phi_1).
\end{align} 
\end{subequations}
where we have defined $\tilde{\beta}_1=\frac{2}{3\sqrt{3}}$ and $\tilde{\beta}_2=\frac{1}{3\sqrt{3}}$, absorbing the factors of $R^2$ in them for 
simplicity. 

With these definitions, we may write the metric as
\be
ds^2=\tilde\b_2 (2ds^2_{AdS_3}+ds_1^2+ds_2^2+dz_1^2+dz_2^2)+(\mathfrak{e}^1)^2+(\mathfrak{e}^2)^2+(\mathfrak{e}^3)^2
\ee
and the RR-forms as ($vol_2=\frac{2}{\tilde \b_1}\mathfrak{e}^1\wedge \mathfrak{e}^2$, $d\psi+P=\frac{\sqrt{2}}{\sqrt{\tilde
\b_2}}\mathfrak{e}^3$)
\begin{subequations}
\begin{align}
\frac{1}{R^2}F_3=&\frac{\tau}{6\sqrt{6}}dz_2 \wedge vol_1-\frac{\tau}{\sqrt{2}}dz_2\wedge \mathfrak{e}^1\wedge \mathfrak{e}^2\\
\frac{1}{R^4}F_5=&\frac{1}{27}\left\{ vol_{AdS_3}\wedge \left[4dz_1\wedge dz_2
+\frac{\tau^2}{2}\left(vol_1+\frac{2}{\tilde \b_1}\mathfrak{e}^1\wedge \mathfrak{e}^2\right)\right]\phantom{\frac{1}{1}}\right.\nonumber\\
&\left.+\frac{\sqrt{2}}{\sqrt{\tilde\b_2}}\mathfrak{e}^3
\wedge \left[  vol_1\wedge \frac{2}{\tilde \b_1}\mathfrak{e}^1\wedge \mathfrak{e}^2
+\frac{\tau^2}{8} dz_1\wedge dz_2\wedge \left(vol_1+\frac{2}{\tilde \b_1}\mathfrak{e}^1\wedge \mathfrak{e}^2\right)\right]\right\},
\end{align}
or as
\begin{align}
F_3=&G_3^{(0)}+ G_1^{12}\wedge\mathfrak{e}^1\wedge \mathfrak{e}^2\label{f3nonab}\\
F_5=&G_5^{(0)}+G_4^3\wedge \mathfrak{e}^3+G_3^{12}\wedge\mathfrak{e}^1\wedge 
\mathfrak{e}^2+G_2^{(3)} \wedge\mathfrak{e}^1\wedge \mathfrak{e}^2\wedge\mathfrak{e}^3\;,\label{f5nonab}
\end{align}
\end{subequations}
where
\begin{align}
\frac{1}{R^4}G_5^{(0)}&=\frac{1}{27} vol_{AdS_3}\wedge \left[4dz_1\wedge dz_2+\frac{\tau^2}{2}vol_1\right]\nonumber\\
\frac{1}{R^4}G_{4\phantom{()}}^3 &=\frac{\sqrt{2}\tau^2}{216\tilde{\beta}^{1/2}_2} dz_1\wedge dz_2\wedge vol_1,\nonumber\\
\frac{1}{R^2}G_3^{(0)}&=\frac{\tau}{6\sqrt{6}}dz_2 \wedge vol_1,\quad \quad \frac{1}{R^4}G_3^{12}=\frac{\tau^2}{27\tilde \b_1}vol_{AdS_3}\nonumber\\
\frac{1}{R^4}G_2^{(3)}&=\frac{4}{27\sqrt{2}}\frac{1}{\tilde{\beta}_2^{1/2}\tilde\b_1}\left(vol_1+\frac{\tau^2}{8}dz_1\wedge dz_2\right)\nonumber\\
\frac{1}{R^2}G_1^{12}&=-\frac{\tau}{\sqrt{2}}dz_2.\label{Gpis}
\end{align}
The matrix $M$ is given by $M_{ij}=g_{ij}+b_{ij}+\a'\epsilon_{ijk}\hat v_k$, so (after absorbing $\a'$ factors in $\hat v_i$)
\begin{equation}
M=\begin{pmatrix}
\tilde{\beta}_1 & \frac{\tau z_1}{\sqrt{2}}\tilde{\beta}_2+\hat{v}_3\quad & -\hat{v}_2 \\ 
-\frac{\tau z_1}{\sqrt{2}}\tilde{\beta}_2-\hat{v}_3 & \tilde{\beta}_1 & \hat{v}_1 \\ 
\hat{v}_2 & -\hat{v}_1 & \tilde{\beta}_2
\end{pmatrix},
\end{equation}

As before, we consider the gauge fixing $\theta=\phi=v_2=0$, so that $\hat{v}=(\cos \psi v_1, \sin\psi v_1, v_3)$, and for simplicity 
we define $\tilde{v}_3=\frac{\tau z_1}{\sqrt{2}}\tilde{\beta}_2+\hat{v}_3$, in such a way that the inverse of $M$ is (\ref{invmatrix}), 
with the replacement $v_3\leadsto \tilde{v}_3$, that is,
\begin{equation}
(M^{-1})^T=\frac{1}{\widetilde{\Delta}}\begin{pmatrix}
\tilde{\beta}_1\tilde{\beta}_2+v_1^2\cos^2\psi\quad &\tilde{v}_3\tilde{\beta}_2+v_1^2\cos\psi\sin\psi\quad & v_1\tilde{v}_3\cos\psi-v_1\tilde{\beta}_1\sin\psi \\ 
-\tilde{v}_3\tilde{\beta}_2+v_1^2\cos\psi\sin\psi & \tilde{\beta}_1\tilde{\beta}_2+v_1^2\sin^2\psi & v_1\tilde{\beta}_1\cos\psi+v_1\tilde{v}_3\sin\psi \\ 
v_1\tilde{v}_3\cos\psi+v_1\tilde{\beta}_1\sin\psi & -v_1\tilde{\beta}_1\cos\psi+v_1\tilde{v}_3\sin\psi & \tilde{v}_3^2+\tilde{\beta}_1^2
\end{pmatrix},
\end{equation}
where the determinant $\det M$ is $\widetilde{\Delta}\equiv\det M=(\tilde{\beta}_1^2+\tilde{v}_3^2)\tilde{\beta}_2+v_1^2\tilde{\beta}_1$.

Under these definitions, we must apply the duality on the following fields\footnote{Note that since the dependence on the angular 
coordinates $(\phi_2,\theta_2)$ is encapsulated into the Maurer-Cartan forms $L_i$, in what follows the subscript $(\phi,\theta)$ 
refers logically to $(\phi_1,\theta_1)$.}
$$
\begin{array}{|l|l|}
  \hline
  Q_{\phi\phi }=G_{\phi\phi}=\tilde\b_2\left(\sin^2\theta_1+\frac{1}{2}\cos^2\theta_1\right) & Q_{\phi 3}=G_{\phi 3}=Q_{3\phi}
  =\frac{\sqrt{2}}{2}\tilde{\beta}_2\cos\theta_1\\
  \hline
  Q_{\theta\theta}=G_{\theta\theta}=\tilde\b_2&\\
  \hline \hline
  Q_{\theta\phi}=B_{\theta\phi}=-\frac{\tau}{2\sqrt{2}}z_1\tilde{\beta}_2\sin\theta_1 &   E_{12}=b_{12}=\frac{2\tau}{\sqrt{2}}\tilde{\beta}_2 z_1\\
  \hline \hline
  E_{11}=E_{22}=g_{11}=\tilde{\beta}_1 &   E_{33}=g_{33}=\tilde{\beta}_2 \\
  \hline
 \end{array}
$$
Using these results and the same procedure as in section 3, we find that the dual metric, dilaton and B field are
\bea
d\hat{s}^2_{AdS_3\times \mathbb{R}^2\times S^2\times S^3}&=&\tilde\b_2\left(2ds^2_{AdS_3}+dz_1^2+dz_2^2+ds_1^2\right)
+\frac{1}{2\widetilde{\Delta}}\tilde{\beta}_1\tilde{\beta}_2 v_1^2(d\psi+\cos\theta_1 d\phi_1)^2\cr
&+&\frac{1}{2\widetilde{\Delta}}\left[(\tilde{\beta}_1\tilde{\beta}_2+v_1^2)dv_1^2+(\tilde{v}_3^2+\tilde{\beta}_1^2)dv_3^2+2v_1\tilde{v}_3 dv_1 dv_3\right]
\label{dualads3metric}\\
\hat\phi&=&\phi-\frac{1}{2}\ln\frac{\tilde\Delta}{\a'^3}\cr
\widehat B&=&-\frac{v_1}{2\tilde\Delta}(\tilde v_3\tilde\b_2dv_1-v_1\tilde\b_1d\hat v_3)\wedge d\psi\cr
&&-\frac{\tilde \b_2 \tau z_1}{2\sqrt{2}}\sin\theta_1 d\theta_1\wedge d\phi_1
+\frac{\tilde \b_2}{2\tilde\Delta}v_1\tilde v_3 \cos\theta_1 d\phi_1\wedge dv_1\cr
&&+\frac{\tilde \b_2}{2\tilde\Delta}\cos\theta_1(\tilde v_3^2+\tilde \b_1^2)d\phi_1\wedge d\hat v_3\cr
&=& -\frac{\tau R z_1}{6\sqrt{6}}vol_1+\frac{\tilde{\b}_2}{2\tilde{\Delta}}\sigma_{\hat{3}}\wedge(v_1\tilde{v}_3 dv_1+(\tilde{v}_3^2+\tilde{\b}_1^2)dv_3)
\label{dualads3bfield}.
\eea
For later use, the $\sqrt{\det g_{int}}$ for this metric ($g_{int}$ is as before the internal, i.e. non-AdS, part of the metric) is 
\be
\sqrt{\det g_{int}}=\a'^3\frac{\sin\theta_1}{2\sqrt{2}}\frac{\tilde\b_1\tilde\b_2^{5/2}}{\tilde \Delta}v_1\label{detads3}.
\ee

With $F_3$ and $F_5$ written as in (\ref{f3nonab}) and (\ref{f5nonab}), we can apply the formulas in the appendix,
reintroduce the factors of $\a'$ in (\ref{Gpis}), (\ref{dualads3metric}) and (\ref{dualads3bfield}) and obtain
the RR-sector T-dual forms $\widehat F_1=\widehat F_3=\widehat F_5=0$ and ($\hat F_6$ and $\hat F_8$ would be redundant, as we 
consider their Poincar\'{e} duals $\hat F_4$ and $\hat F_2$)
\bea
\widehat{F}_2&=&e^{\phi-\hat{\phi}}\left\{-A_0 G_2^{(3)}+G_1^{12}\wedge(A_2\hat{\mathfrak{e}}^1-A_1\hat{\mathfrak{e}}^2-A_0 \hat{\mathfrak{e}}^3)
\right\}\cr
\widehat F_4&=&e^{\phi-\hat \phi}\left\{A_3 G_4^{3}+G_3^{12}\wedge(A_2\hat{\mathfrak{e}}^1-A_1\hat{\mathfrak{e}}^2-A_0 \hat{\mathfrak{e}}^3)
+G_3^{(0)}(A_1 \hat{\mathfrak{e}}^1+A_2\hat{\mathfrak{e}}^2+A_3\hat{\mathfrak{e}}^3)\right.\cr
&&\left.+ G_2^{(3)}\wedge (A_3\hat{\mathfrak{e}}^1\wedge \hat{\mathfrak{e}}^2+A_1\hat{\mathfrak{e}}^2\wedge \hat{\mathfrak{e}}^3
+A_2\hat{\mathfrak{e}}^3\wedge \hat{\mathfrak{e}}^1)
+A_3 G_1^{12}\hat{\mathfrak{e}}^1\wedge \hat{\mathfrak{e}}^2\wedge
\hat{\mathfrak{e}}^3\right\}\;,\label{f2f4}
\eea
where as before, $e^{\phi-\hat\phi}=\sqrt{\tilde\Delta}\a'^{-3/2}$, $\a'^{3/2}e^{\phi-\hat\phi}A_0=\tilde \b_1\sqrt{\tilde\b_2}$ and 
$\a'^{3/2}e^{\phi-\hat\phi}A_a={\cal A}_a$, and the dual vielbeins are
\begin{subequations}
\begin{align}
\hat{\mathfrak{e}}^{\prime 1}_{AdS_3}&=-\frac{\tilde{\beta}_1^{1/2}}{\sqrt{2}\tilde{\Delta}}\left[(\tilde{\beta}_1\tilde{\beta}_2
+v_1^2)dv_1+v_1 \tilde{v}_3 dv_3+v_1 \tilde{v}_3\tilde{\beta}_2(d\psi+\cos \theta_1 d\phi_1)\right]\\
\hat{\mathfrak{e}}^{\prime 2}_{AdS_3}&=\frac{\tilde{\beta}_1^{1/2}}{\sqrt{2}\tilde{\Delta}}\left[\tilde{\beta}_2\tilde{v}_3dv_1
-v_1 \tilde{\beta}_1 dv_3-v_1 \tilde{\beta}_1\tilde{\beta}_2(d\psi+\cos \theta_1 d\phi_1)\right]\\
\hat{\mathfrak{e}}^3_{AdS_3}&=-\frac{\tilde{\beta}_2^{1/2}}{\sqrt{2}\tilde{\Delta}}\left[v_1 \tilde{v}_3 dv_1
+(\tilde{v}_3^2+\tilde{\beta}_1^2)dv_3-v_1^2 \tilde{\beta}_1 (d\psi+\cos \theta_1 d\phi_1)\right].
\end{align}
\end{subequations}

\subsection{Domain Wall and its non-abelian T-dual}

The  Domain Wall solution which has as limits the above $AdS_3$ and $AdS_5$ solution is given by 
\begin{subequations}
\begin{align}
\frac{1}{R^2}ds^2_{DW}=&e^{2A}(-dt^2+dx^2)+e^{2B}(dx_1^2+dx_2^2)+d\rho^2\nonumber\\
&\quad+\frac{1}{6} e^{2U}(ds_1^2+ds_2^2)+\frac{1}{9}e^{2V}(\sqrt{2}L_3+\cos\theta_1 d\phi_1)^2\\
\frac{1}{R^2}B=&\frac{-\tau}{6}x_1 (vol_1-vol_2)\\
\frac{1}{R^2}F_3=&\frac{\tau}{6}dx_2\wedge (vol_1-vol_2)\\
\frac{1}{R^4}F_5=&4 e^{2A+2B-V-4U}dt\wedge dx\wedge dx_1\wedge dx_2\wedge d\rho+\frac{1}{27} 
(\sqrt{2}L_3+\cos\theta_1 d\phi_1)\wedge vol_1\wedge vol_2\nonumber\\
&+\frac{\tau^2}{36} dx_1\wedge dx_2\wedge (\sqrt{2}L_3+\cos\theta_1 d\phi_1)\wedge (vol_1+vol_2)\\
&+ \frac{\tau^2}{12}e^{2A-2B-V}dt\wedge dx\wedge d\rho \wedge(vol_1+vol_2).\nonumber
\end{align}
\end{subequations}
Here $\tau$ is a constant and $A, B, U, V$ are functions of the radial coordinate $\rho$. From this solution, we see that we can 
recover $AdS_5\times T^{(1,1)}$ by setting the constant $\tau=0$ and $A=B=\rho$ and $U=V=0$. On the other hand, to recover the 
$AdS_3\times \mathbb{R}^2\times S^2\times S^3 $ solution, we set
\begin{equation}
A=\frac{3^{3/4}}{\sqrt{2}}\rho,\qquad B=U=-V=\frac{1}{4}\ln\left(\frac{4}{3}\right),
\end{equation}
and change variables by $x_i\leadsto z_i/\sqrt{6}$\ .

As before, the T-dual model is given by 
\bea
d\hat{s}^2_{DW}&=&R^2e^{2A}(-dt^2+dx^2)+R^2e^{2B}(dx_1^2+dx_2^2)+R^2d\rho^2\cr
&&+\frac{R^2}{6} e^{2U}ds_1^2+\frac{1}{2\bar{\Delta}}\bar{\beta}_1\bar{\beta}_2 v_1^2(d\psi+\cos\theta_1 d\phi_1)^2\cr
&&+\frac{1}{2\bar{\Delta}}\left\{(\bar{\beta}_1\bar{\beta}_2+v_1^2)dv^2_1+(\bar{\beta}_1^2+\bar{v}_3^2)dv_3^2+2v_1\bar{v}_3 dv_1 dv_3\right\}\cr
\widehat{B}&=& -\frac{\tau R x_1}{6}vol_1+\frac{\bar{\b}_2}{2\bar{\Delta}}\sigma_{\hat{3}}\wedge(v_1\bar{v}_3 dv_1+(\bar{v}_3^2+\bar{\b}_1^2)dv_3)\cr
\hat\phi&=&\phi-\frac{1}{2}\ln\frac{\bar\Delta}{\a'^3}\;,
\eea
where we have defined
\begin{equation}
\bar{\beta}_1=\frac{1}{3}e^{2U},\quad \bar{\beta}_2=\frac{2}{9}e^{2V}, \quad \bar{v}_3=\frac{\tau}{3}x_1+\hat{v}_3,\;\;\;
\bar\Delta=(\bar\b_1^2+\bar v_3^2)\bar\b_2+v_1^2\bar\b_1\;,
\end{equation}
and as before we absorbed $R^2$ factors in $\bar\b_i$ and $\a'$ in $v_i$.

We can easily see that we can obtain the correct limits in the NS-NS sector. The UV and IR limits of the T-dual solution to the domain wall are 
the non-abelian T-duals of the $AdS_5\times T^{(1,1)}$ and the $AdS_3\times \mathbb{R}^2\times S^2\times S^3 $ solutions, respectively. 

In the RR sector, we could verify term by term that the equality holds, but alternatively, one can find the  
 RR-forms components in the same way as in (\ref{Gpis}). In the present case, we obtain
\begin{align}
\frac{1}{R^4}G_5^{(0)}&=dt\wedge dx\wedge d\rho\wedge \left(4e^{2A+2B-V-4U} dx_1\wedge dx_2+\frac{\tau^2}{12}e^{2A-2B-V}vol_1\right)\\
\frac{1}{R^4}G_{4\phantom{()}}^3&=\frac{\sqrt{2}\tau^2}{36 \bar{\b}^{1/2}_2} dx_1\wedge dx_2 \wedge vol_1\\
\frac{1}{R^2}G^{(0)}_3&=\frac{\tau}{6}dx_2\wedge vol_1\; ,\quad \frac{1}{R^2}G_3^{12}=\frac{\tau^2 }{6\bar{\b}_1}e^{2A-2B-V}dt\wedge dx\wedge d\rho\\
\frac{1}{R^4}G_2^{(3)}&= \frac{2\sqrt{2}}{27 \bar{\b}_1 \bar{\b}_2^{1/2}}vol_1+\frac{2\sqrt{2}\tau^2}{36 \bar{\b}_1 \bar{\b}_2^{1/2}}dx_1\wedge dx_2\\
\frac{1}{R^2}G_1^{12}&=-\frac{\tau }{3\bar{\b}_1} dx_2\; ,
\end{align}

Then the T-dual RR-forms are as in (\ref{f2f4}), i.e.,
\bea
\widehat{F}_2&=&e^{-\hat{\phi}}\left\{-A_0 G_2^{(3)}+G_1^{12}\wedge(A_2\hat{\mathfrak{e}}^1-A_1\hat{\mathfrak{e}}^2-A_0 \hat{\mathfrak{e}}^3)
\right\}\cr
\widehat F_4&=&e^{-\hat \phi}\left\{A_3 G_4^{3}+G_3^{12}\wedge(A_2\hat{\mathfrak{e}}^1-A_1\hat{\mathfrak{e}}^2-A_0 \hat{\mathfrak{e}}^3)+G_3^{(0)}\wedge(A_1\hat{\mathfrak{e}}^1+A_2\hat{\mathfrak{e}}^2+A_3\hat{\mathfrak{e}}^3)
\right.\cr
&&\left.+ G_2^{(3)}\wedge(A_3 \hat{\mathfrak{e}}^1\wedge \hat{\mathfrak{e}}^2+A_2 \hat{\mathfrak{e}}^3\wedge \hat{\mathfrak{e}}^1+A_1 \hat{\mathfrak{e}}^2\wedge \hat{\mathfrak{e}}^3)+A_3 G_1^{12}\hat{\mathfrak{e}}^1\wedge \hat{\mathfrak{e}}^2\wedge
\hat{\mathfrak{e}}^3\right\}\;.
\eea
Finally, we can also compute the vielbeins and see that they have the 
correct limits, therefore the RR-sector also has the correct limits. For instance, the frame field $\mathfrak{e}^3$ of the Domain Wall is
\begin{subequations}
\begin{equation}
\mathfrak{e}^3_{AdS(DW)}=-\frac{\bar{\beta}_2^{1/2}}{\sqrt{2}\bar\Delta}\left[v_1 \bar{v}_3 dv_1
+(\bar{v}_3^2+\bar{\beta}_1^2)dv_3-v_1^2 \bar{\beta}_1 (d\psi-\cos \theta_1 d\phi_1)\right],
\end{equation} 
and we can easily verify that the UV and IR limits are the frame field $\mathfrak{e}^3$ in the AdS${}_5$, AdS${}_3$
\begin{align}
\mathfrak{e}^3_{AdS_5}&=-\frac{\beta_2^{1/2}}{\sqrt{2}\bar\Delta}\left[v_1 v_3 dv_1+(v_3^2+\beta_1^2)dv_3-v_1^2 \beta_1
(d\psi+\cos \theta_1 d\phi_1)\right]\\
\mathfrak{e}^3_{AdS_3}&=-\frac{\tilde{\beta}_2^{1/2}}{\sqrt{2}\bar\Delta}\left[v_1 \tilde{v}_3 dv_1
+(\tilde{v}_3^2+\tilde{\beta}_1^2)dv_3-v_1^2 \tilde{\beta}_1 (d\psi+\cos \theta_1 d\phi_1)\right]
\end{align}
\end{subequations}
respectively.

\section{Dual conformal field theories, central charges and RG flow}

An interesting question is, what happens to the conformal field theories dual to the gravity backgrounds with AdS factor under nonabelian T-duality
on the extra dimensional space?
The answer is not obvious. Abelian T-duality on a direction transverse to a D$p$-brane turns it into a D$(p+1)$-brane, but if the original direction is 
infinite in extent, the T-dual direction is infinitesimal in extent. However, this discussion makes sense only in the region far from the 
region where AdS/CFT is relevant, the core of the D-brane. 

Naively, abelian T-duality on the transverse part of a gravity dual should increase the dimensionality of the brane, therefore of the field 
theory dual to the background. But if we perform a nonabelian T-duality on a space with an AdS factor, in such a way that the AdS factor is not 
affected, and moreover the T-duality does not introduce a new AdS direction, then it seems that the dimensionality of the dual conformal field theory 
is unaffected. And yet since the gravity dual is modified, it is logical to assume that the conformal field theory is modified as well. 

To understand the effect of nonabelian T-duality on the conformal field theory, we need some probes of the transverse space in AdS/CFT. Such probes 
are for instance wrapped branes, dual to solitonic states in the field theory, like the example of the 5-brane wrapped on $S^5$ in $AdS_5\times S^5$, 
giving the baryon vertex operator \cite{Witten:1998xy}.\footnote{Baryon vertex probes in this context, but in other dimensions have been considered in 
\cite{Lozano:2013oma} and \cite{Lozano:2014ata}.} 
But a more relevant probe was considered in \cite{Macpherson2014}, namely the central charge
of the dual field theory as a function of the number of branes. 

One can calculate Page charges in a gravitational background, and identify those with the number of branes that generate the geometry.
For the central charge of the dual conformal field theory, a slight generalization of the usual formula was provided in \cite{Macpherson2014}. For a metric
on $M^D=AdS_{d+2}\times X^n$, of the type
\be
ds_D^2=A\; d\vec{z}_{(1,d)}^2+AB\; dr^2+g_{ij}d\theta^id\theta^j\;,
\ee
with a dilaton $\phi$, define the modified internal volume as 
\be
\hat V_{int}=\int d\vec{\theta}\sqrt{e^{-4\phi}\det[g_{int}]A^d}\;
\ee
and then $\hat H=\hat V_{int}^2$. Then the central charge is given by
\be
{\cal C}=d^d\frac{B^{d/2}\hat H^{\frac{2d+1}{2}}}{G_N(\hat H')^d}\;
\ee
where $G_N=(\a')^{\frac{D}{2}-1}$ is the Newton constant in $D$ dimensions and prime denotes the derivative with respect to $r$.

The expectation of increase in dimensionality through T-duality affects the D-brane charges of the gravity background. For a geometry with 
an $AdS_5$ factor in type IIB, generated only by D3-branes (with only D3-brane Page charges), after T-duality we expect the geometry to 
be generated by D4- and D6-branes only, i.e. to have only D4- and D6-brane Page charges
\bea
\mathcal{Q}^{Page}_{D4}&=&\frac{1}{2\kappa_{10}^2T_{D4}}\int_{\Sigma_4}(\hat F_4-\hat B\wedge \hat F_2)\cr
\mathcal{Q}^{Page}_{D6}&=&\frac{1}{2\kappa_{10}^2T_{D6}}\int_{\Sigma_2}\hat F_2.
\eea
For an abelian T-duality, we would expect only D4-brane charge, but for nonabelian T-duality (in some sense a T-duality on 3 coordinates), 
the expectation, confirmed by a calculation, is that only D6-brane charges appear. One can calculate the central charges and express them as a 
function of the Page charges. In the $AdS_5\times S^5$ case, we find that ${\cal C}=32\pi^3 R^8\a'^{-4}=2\pi^5 N_{D3}^2$ before, and 
${\cal C}=(8\pi^5/3) R^8\a'^{-4}=(2\pi^5/24)N^2_{D6}$ after the nonabelian T-duality, leading to the relation\footnote{The formula in \cite{Macpherson2014}
is actually with a factor of 3 instead of 24, since different conventions for T-duality were considered, with $L_i=\sigma_i$ instead of $L_i=\sigma_i/\sqrt{2}$, 
giving an extra $2\sqrt{2}$ in the quantization of the Page charges after T-duality.}
\be
\frac{{\cal C}_{before}}{{\cal C}_{after}}=\frac{24N^2_{D3}}{N^2_{D6}}\;,\label{centrallaw}
\ee
which is found to be satisfied also in other cases of non-abelian T-duality on type IIB geometries generated by D3-branes.

An interesting question which we will try to answer in this section is whether a similar formula is valid in more general contexts in the case 
of geometries with an AdS factor.

\subsection{Page charges}

\begin{itemize}

\item In the case of section 3, the starting geometry is in type IIA, the reverse of the situation considered in \cite{Macpherson2014}. 
Since $F_2=0$ in the background before T-duality, $\mathcal{Q}^{Page}_{D6}=0$, and we only have a nonzero result for
\bea
N_{D4}=|\mathcal{Q}^{Page}_{D4}|&=&\frac{R^3}{2\kappa_{10}^2T_{D4}}\int_{y_1}^{y_2}\eta(y)dy\int_{X_3}vol(S^2)\wedge L_3\cr
&=&\left(\frac{R}{2\pi \sqrt{\a'}}\right)^3\frac{2\sqrt{2}}{9}(y_2-y_1)\left(1-c\frac{y_1+y_2}{2}\right)4\pi^2\sqrt{2}\cr
&\equiv&\left(\frac{R}{l_s}\right)^3\frac{2}{9\pi}K.
\label{ND4}
\eea
After the nonabelian T-duality, we have calculated in section 3 that $\mathcal{Q}^{Page}_{D3}=\mathcal{Q}^{Page}_{D5}=0$ and
\bea
N_{D7}=|\mathcal{Q}^{Page}_{D7}|&=&\frac{R^3}{\a'^{3/2}}\frac{4\sqrt{2}}{9}(y_2-y_1)\left(1-c \frac{y_1+y_2}{2}\right)\cr
&=&\left(\frac{R}{l_s}\right)^3\frac{4\sqrt{2}}{9}K.\label{ND6}
\eea

\item In the case of section 4, the we have a Domain Wall solution that interpolates between an $AdS_5\times T^{1,1}$ and an 
$AdS_3\times \mathbb{R}^2\times S^2\times S^3$. This can be also found in the ${\cal N}=4$ D=5 gauged supergravity arising as a consistent
KK truncation of type IIB on $T^{1,1}$ \cite{Gauntlett2014}, and as such it can be interpreted as an RG flow between two fixed points in the dual field 
theory. A relevant question is then, is the ratio of the central charges before and after the nonabelian T-duality modified by the RG flow?

For $AdS_5\times T^{1,1}$, the Page charges before and after the nonabelian T-duality were found in \cite{Macpherson2014}, 
$\mathcal{Q}^{Page}_{D5}=\mathcal{Q}^{Page}_{D7}=0$ and $|\mathcal{Q}^{Page}_{D3}|=N_{D3}$ before, and 
$|\mathcal{Q}^{Page}_{D6}|=N_{D6}$, $\mathcal{Q}^{Page}_{D4}=0$ after the T-duality, with (in our conventions)
\begin{equation}
N_{D3}=\frac{4 R^4}{27 \pi \alpha^{\prime 2 }},\quad N_{D6}=\frac{4\sqrt{2}}{27 \alpha^{\prime 2}}R^4.
\end{equation} 

For $AdS_3\times \mathbb{R}^2\times S^2\times S^3$, the Page charges before the T-duality were found in \cite{Gauntlett2014}. Assuming that
$\mathbb{R}^2$ is compactified to a $T^2=S^1_{(1)}\times S^1_{(2)}$ 
with period $2\pi Rd_i\sqrt{6}$, and defining $s(S)$ as a homology 2-cycle generator in $S^2\times S^3$, 
one has the integers
\bea
Q_{N5}&=&\frac{1}{(2\pi l_s)^2}\int_{S^1_{(1)}\times s(S)}H\cr
Q_{D5}&=&\frac{1}{(2\pi l_s)^2}\int_{S^1_{(2)}\times s(S)}dC_2\;
\eea
and the (D3-brane)  Page charge quantization condition is 
\be
\frac{1}{(2\pi l_s)^4}\int_{\Sigma_5}(F_5-B\wedge dC_2)\in \mathbb{Z}.
\ee
For $\Sigma_5=S^2\times S^3$, one obtains an integer 
\be
N=\left(\frac{R}{l_s}\right)^4\frac{vol(T^{1,1})}{4\pi^4}\;, 
\ee
and for $\Sigma_5=T^2\times M_3$, where $M_3$ is a homology 3-cycle
generator in $S^2\times S^3$ , one obtains an integer 
\be
\bar N=\left(\frac{R}{l_s}\right)^4\frac{8 d_1 d_2}{9}=-\frac{1}{2}Q_{N5}Q_{D5}. 
\ee
Moreover, the above flux quantization is actually valid over the whole domain wall solution. 

After the T-duality, we have $F_2$ and $F_4$, so we need to consider the quantization of D4-brane Page charges
\be
\frac{1}{(2\pi l_s)^3}\int_{\Sigma_4}(F_4-B\wedge F_2)\in \mathbb{Z}
\ee
and
\be
\frac{1}{(2\pi l_s)}\int_{\Sigma_2} F_2\in \mathbb{Z}.
\ee
For $\Sigma_2=T^2$, we obtain 
\be
N_{D6}=-\frac{\tau^22\sqrt{2}}{216}4\pi^2 \frac{6d_1d_2}{2\pi l_s}\frac{R^4}{l_s^3}\;,
\ee
and for $\Sigma_2=S^2$, we obtain 
\be
\bar N_{D6}=-\frac{2\sqrt{2}}{27}\frac{4\pi}{2\pi l_s}\frac{R^4}{l_s^3}.
\ee

\end{itemize}

\subsection{Central charges}

\begin{itemize}

\item For the case in section 3, the central charge before the T-duality is obtained using  $A=R^2 r^2$, $B=r^{-4}$ and $d=3$, leading to 
($\int L_1\wedge L_2\wedge L_3=2\pi^2\sqrt{2}$)
\be
\hat V_{int}=\frac{\a'r^3 R^6}{l}(2\pi)\frac{4\pi^2}{9}(y_2-y_1)\left(1-\frac{c(y_1+y_2)}{2}\right)\equiv \frac{\a'r^3 R^6}{l}\frac{8\pi^3}{9} K\;,
\ee
and therefore
\be
{\cal C}_{before}=\frac{R^6}{8\a'^3}\frac{8\pi^3}{9}\frac{K}{l}\;,
\ee
where the Page charge quantization condition (\ref{ND4}) means that we can write $R^3/\a'^{3/2}$ as a function of $N_{D4}$, giving
\be
{\cal C}_{before}=\frac{9\pi^5}{4}\frac{ N_{D4}^2}{Kl}.
\ee

After the T-duality, the central charge is found using the same $A=R^2 r^2$, $B=r^{-4}$ and $d=3$, leading to
(also using the $\sqrt{\det g_{int}}$ calculated in (\ref{detads5}))
\be
\hat V_{int}=\frac{\a'R^6 r^3}{2l}(2\pi)^2\frac{K}{9}\int \frac{dv_1}{\a'}\frac{v_1}{\a'}\int \frac{dv_3}{\a'}.
\ee
To calculate the integral over the $v_i$, we can use as another gauge fixing, related to the previous coordinates by 
$v_1/\a'\leadsto\rho \cos \chi$ and $v_3/\a'\leadsto\rho \sin \chi$ with $\rho,\chi\in[0,\pi]$, leading to a value of $2\pi^3/3$ for the integral.\footnote{The
rangle of $\rho$ was defined in \cite{Lozano:2014ata}, and was then used for the calculation of central charges in \cite{Macpherson2014}.}
We then obtain
\be
{\cal C}_{after}=\frac{\pi^5K}{54l}\left(\frac{R}{l_s}\right)^6\;,
\ee
and from the Page charge quantization condition (\ref{ND6}) we can write $R^3/\a'^{3/2}$ as a function of $N_{D7}$ , giving
\be
{\cal C}_{after}=\frac{3\pi^5}{64 Kl}N_{D7}^2
\ee

We see that the ratio is
\be
\frac{{\cal C}_{before}}{{\cal C}_{after}}=\frac{48N_{D4}^2}{N_{D7}^2}\;,
\ee
which is basically the same as in (\ref{centrallaw}), with the obvious generalization to $N^2_{Dp}/N^2_{Dp+3}$, and an extra factor of 2 which is 
probably the effect of a different normalization.

\item For the case in section 4, on the $AdS_5\times T^{1,1}$ side, the central charge before the T-duality was found to be \cite{Macpherson2014}
\be
{\cal C}^{(1)}_{before}=\frac{\pi^3 R^8}{27 \a'^4}=\frac{27}{8}\pi^5 N^2_{D3}\;,
\ee
and after the T-duality
\be
{\cal C}^{(1)}_{after}=\frac{2R^8\pi^5\lambda\lambda_0^4}{3\a'^4}=\frac{9}{64}\pi^5 N^2_{D6}\;,
\ee
leading to the ratio in (\ref{centrallaw}).
On the $AdS_3\times T^2\times S^2\times S^3$ side, the central charge before the T-duality is \cite{Gauntlett2014}
\bea
{\cal C}^{(2)}_{before}&=&\frac{3R_{AdS_3}}{2G_3}=\frac{3}{2}\left(\frac{R}{l_s}\right)^8\frac{8d_1d_2}{9}\frac{vol(T^{1,1})}{4\pi^4}\cr
&=&\frac{3}{2}|N Q_{N5}Q_{D5}|=3|N\bar N|=3N_{D3}\bar N_{D3}.
\eea
Here $G_3$ is the effective Newton's constant, obtained from the dimensional reduction of the action in string frame, thus proportional to 
$(R/l_s)^7vol(T^{1,1})(2\pi d_1)(2\pi d_2)$.
After the T-duality, using the $\sqrt{\det g_{int}}$ calculated in (\ref{detads3}), and doing the integration over $v_i$ in the same way as 
in the case in section 3, with result $2\pi^3/3$, we obtain 
\be
\hat V_{int}=\frac{R^8}{r}\frac{12(2\pi)^4d_1d_2}{\sqrt{2}}\left(\frac{1}{3\sqrt{3}}\right)^{7/2}\frac{2\pi^3}{3}\;,
\ee
leading to 
\be
{\cal C}_{after}=\frac{32\sqrt{2}\pi^7}{3^{21/4}}d_1d_2\frac{R^8}{l_s^8}=\frac{4}{\tau^2}3^{-1/4}\sqrt{2}\pi^6N_{D6}\bar N_{D6}.
\ee
The ratio of central charges before and after the T-duality can therefore be expressed as 
\be
\frac{{\cal C}_{before}^{(2)}}{{\cal C}_{after}^{(2)}}=\frac{3^{5/4}\sqrt{2}\tau^2}{8\pi^6}\frac{N_{D3}\bar N_{D3}}{N_{D6}\bar N_{D6}}.
\ee
Note that now we can fix $\tau$ such that the prefactor equals 24, obtaining 
\be
\frac{{\cal C}_{before}^{(2)}}{{\cal C}_{after}^{(2)}}=\frac{24N_{D3}\bar N_{D3}}{N_{D6}\bar N_{D6}}\;,
\ee
which is essentially the same formula (\ref{centrallaw}) that was valid on the $AdS_5$ side of the domain wall. The factor $\tau$ is related to a
redefinition of the fields, coupled to a rescaling of the $x_i$ (or $z_i$) coordinates \cite{Gauntlett2014}, 
which are the two coordinates that change from the $AdS_5$ on one side of the domain wall to 
a $AdS_3\times T^2$ on the other. It is therefore not surprising that changing $\tau$ allows us to change the normalization of the central charge 
dual to $AdS_3$, with respect to the one dual to $AdS_5$.

\end{itemize}

\section{Conclusions}

In this paper we have studied the nonabelian T-duals of some backgrounds with ${\cal N}=1$ supersymmetry and an AdS factor, that 
can have an AdS/CFT interpretation. We have considered the nonabelian T-dual of a type IIA solution with an $AdS_5$ factor, giving a 
type IIB solution with an $AdS_5$ factor, and the nonabelian T-dual of a type IIB domain wall solution that interpolates between 
$AdS_5\times T^{1,1}$ and $AdS_3\times \mathbb{R}^2\times S^2\times S^3$. 

We have probed the interpretation of nonabelian T-duality of these solutions from the point of view of the dual conformal field theory through a 
calculation of the central charges. We have found that the simple law (\ref{centrallaw})  found in  \cite{Macpherson2014}
for the ratio of central charges before and after the T-duality holds in all cases, with the obvious generalization of $N_{D3}^2/N_{D6}^2$
to $N_{Dp}^2/N_{Dp+3}^2$ or to $N_{D3}\bar N_{D3}/N_{D6}\bar N_{D6}$.
In the case of the type IIB dowain wall solution, we obtained the usual $\propto N^2$ behaviour, and 
on the $AdS_3$ side we could fix the normalization of the central charge by using a rescaling parameter $\tau$, in order to obtain the same 
law (\ref{centrallaw}) valid on the $AdS_5$ side of the domain wall. In order to
understand better the effect of nonabelian T-duality on gravity duals with AdS factors, one needs to study also 
other probes of the geometry, but we leave this for future work.

{\bf Acknowledgements}. We would like to thank Carlos N\'{u}\~nez for discussions, and Ozgur Kelekci, Yolanda Lozano, Niall Macpherson and 
Eoin O'Colgainn for comments and for pointing out to us references that we had missed in the first version of the paper. 
The research of HN is supported in part by CNPQ grant 301709/2013-0
and FAPESP grant 2013/14152-7 and the work of TA is supported by CNPq grant 140588/2012-4.

\appendix

\section{Nonabelian T-duality action on RR fields}

To act with nonabelian T-duality on the RR fields, one first writes the $p$-form field strengths in the form
\be
F_p=G_p^{(0)}+G_{p-1}^a\wedge \mathfrak{e}^a+\frac{1}{2}G^{ab}_{p-2}\wedge \mathfrak{e}^a\wedge \mathfrak{e}^b+
G_{p-3}^{(3)}\wedge \mathfrak{e}^1\wedge \mathfrak e^2\wedge \mathfrak{e}^3.
\ee
Using a similar decomposition for the T-dual $p$-forms $\hat F_p$ in terms of the T-dual vielbeins $\mathfrak{e}'$,
\be
\hat F_p=\hat G_p^{(0)}+\hat G_{p-1}^a\wedge {\mathfrak{e}'}^a+\frac{1}{2}\hat G^{ab}_{p-2}\wedge {\mathfrak{e}'}^a\wedge {\mathfrak{e}'}^b+
\hat G_{p-3}^{(3)}\wedge {\mathfrak{e}'}^1\wedge {\mathfrak e'}^2\wedge {\mathfrak{e}'}^3\;,
\ee
we have the transformation rules
\bea
\hat G_p^{(0)}&=&e^{\phi-\hat \phi}(-A_0G_p^{(3)})+A_a G_p^a)\cr
\hat G^a_{p-1}&=&e^{\phi-\hat\phi}\left(-\frac{A_0}{2}\epsilon^{abc}G^{bc}_{p-1}+A_b G_{p-1}^{ab}+A_a G_{p-1}^{(0)}\right)\cr
\hat G^{ab}_{p-2}&=&e^{\phi-\hat\phi}\left[\epsilon^{abc}(A_c G_{p-2}^{(3)}+A_0 G^c_{p-2})-(A_a G^b_{p-2}-A_b G^a_{p-2})\right]\cr
\hat G^{(3)}_{p-3}&=&e^{\phi-\hat\phi}\left(\frac{A_a}{2}\epsilon^{abc}G^{bc}_{p-3}+A_0 G_{p-3}^{(0)}\right).
\eea
Here, defining $y_i=b_i+\a'v_i$ as before and
\bea
z^i&=&\frac{y^i}{\sqrt{\det g}}\cr
\zeta^a&=&{\kappa^a}_i z^i={\kappa^a}_i\frac{y^i}{\sqrt{\det g}}\;
\eea
the coefficients of the transformation rules are 
\bea
A_0&=&\frac{1}{\sqrt{1+\zeta^2}}=\frac{\sqrt{\det g}}{\sqrt{\det g+({\kappa^a}_i y^i)^2}}\cr
A_a&=&\frac{\zeta_a}{\sqrt{1+\zeta^2}}=\frac{{\kappa^a}_i y^i}{\sqrt{\det g +({\kappa^a}_i y^i)^2}}.
\eea

\subsection{Particular cases for the coefficients}

In the case of section 3, we have $b_i=0$, so $y^i=\a'\hat v^i$, leading to 
\be
{\kappa^a}_i=\a' R{\rm diag}(\sqrt{\b_1}\hat v_1,\sqrt{\b_1}\hat v_2,\sqrt{\b_2}v_3)\;,
\ee
and $\det g=R^6\b_1^2\b_2$ and 
\be
\sqrt{\det g+({\kappa^a_iy^i)^2}}=R\sqrt{\b_2(R^4\b_1^2+\a'^2v_3^2)+\a'^2\b_1 v_1^2}=\sqrt{\Delta}.
\ee
We also have $e^{\phi-\hat \phi}=\sqrt{\Delta}/\a'^{3/2}$, so 
\bea
\a'^{3/2}e^{\phi-\hat\phi}A_a&\equiv& {\cal A}_a=R\a'(\sqrt{\b_1}v_1\cos\psi,\sqrt{\b_1}v_1\sin\psi,\sqrt{\b_2}v_3)\cr
\a'^{3/2}e^{\phi-\hat\phi}A_0&=&R^3\b_1\sqrt{\b_2}.
\eea
In the case of section 4, the same formulas apply, with the replacement of $v_3$ with $\tilde v_3$.


\begin{thebibliography}{99}

\bibitem{Green1987}
M. Green, J. Schwarz and E. Witten, \emph{Superstring Theory Vol. 1: Introduction}, Cambridge University Press, (1988).

\bibitem{Maldacena1997}
  J.~M.~Maldacena,
  ``The Large N limit of superconformal field theories and supergravity,''
  Int.\ J.\ Theor.\ Phys.\  {\bf 38}, 1113 (1999)
  [Adv.\ Theor.\ Math.\ Phys.\  {\bf 2}, 231 (1998)]
  [hep-th/9711200].


\bibitem{Aharony1999a}
  O.~Aharony, S.~S.~Gubser, J.~M.~Maldacena, H.~Ooguri and Y.~Oz,
  ``Large N field theories, string theory and gravity,''
  Phys.\ Rept.\  {\bf 323}, 183 (2000)
  [hep-th/9905111].


\bibitem{Polchinski1998}
J. Polchinski, \emph{String Theory Vol. 1: An Introduction to the Bosonic String}, Cambridge University Press, (1988).

\bibitem{Greene1996}
B.~R.~Greene,  ``\emph{String Theory on Calabi-Yau Manifolds},'' 
\emph{Theoretical Advanced Study Institute in Elementary Particle Physics (TASI 96): Fields, Strings, and Duality}, (1996) [hep-th/9702155].

\bibitem{Strominger1996}
  A.~Strominger, S.~T.~Yau and E.~Zaslow,
  ``Mirror symmetry is T duality,''
  Nucl.\ Phys.\ B {\bf 479}, 243 (1996)
  [hep-th/9606040].

\bibitem{Frenke2005}
E. Frenkel,  ``\emph{Lectures on the Langlands program and conformal field theory},'' 
\emph{Les Houches School of Physics: Frontiers in Number Theory, Physics and Geometry, 
Les Houches, France 9-21 Mar 2003}, \emph{DARPA Workshop on Langlands Program and Physics, 
Princeton, New Jersey 8-10 Mar 2005} [hep-th/0512172].

\bibitem{Frenke2009}
E. Frenkel,  ``\emph{Gauge Theory and Langlands Duality},'' (2009) [math.RT/0906.2747v1].

\bibitem{de la Ossa:1992vc} 
  X.~C.~de la Ossa and F.~Quevedo,
  ``Duality symmetries from nonAbelian isometries in string theory,''
  Nucl.\ Phys.\ B {\bf 403}, 377 (1993)
  [hep-th/9210021].

\bibitem{Sfetsos2010}
  K.~Sfetsos and D.~C.~Thompson,
  ``On non-abelian T-dual geometries with Ramond fluxes,''
  Nucl.\ Phys.\ B {\bf 846}, 21 (2011)
  [arXiv:1012.1320 [hep-th]].


\bibitem{Itsios2013}
  G.~Itsios, C.~Nunez, K.~Sfetsos and D.~C.~Thompson,
  ``Non-Abelian T-duality and the AdS/CFT correspondence:new N=1 backgrounds,''
  Nucl.\ Phys.\ B {\bf 873}, 1 (2013)
  [arXiv:1301.6755 [hep-th]].


\bibitem{Buscher1988} 
  T.~H.~Buscher,
  ``Path Integral Derivation of Quantum Duality in Nonlinear Sigma Models,''
  Phys.\ Lett.\ B {\bf 201}, 466 (1988).
 
\bibitem{Macpherson2014}
  N.~T.~Macpherson, C.~N\'{u}\~nez, L.~A.~Pando Zayas, V.~G.~J.~Rodgers and C.~A.~Whiting,
  ``Type IIB supergravity solutions with AdS$_{5}$ from Abelian and non-Abelian T dualities,''
  JHEP {\bf 1502}, 040 (2015)
  [arXiv:1410.2650 [hep-th]].


\bibitem{Barranco:2013fza} 
  A.~Barranco, J.~Gaillard, N.~T.~Macpherson, C.~N\'{u}\~nez and D.~C.~Thompson,
  JHEP {\bf 1308}, 018 (2013)
  [arXiv:1305.7229, arXiv:1305.7229 [hep-th]].
  
\bibitem{Lozano:2014ata} 
  Y.~Lozano and N.~T.~Macpherson,
  ``{\em A new AdS$_{4}$/CFT$_{3}$ dual with extended SUSY and a spectral flow},''
  JHEP {\bf 1411}, 115 (2014)
  [arXiv:1408.0912 [hep-th]].

\bibitem{Lozano:2012au} 
  Y.~Lozano, E.~\'{O} Colg\'{a}in, D.~Rodrí\'{i}guez-G\'{o}mez and K.~Sfetsos,
  ``Supersymmetric $AdS_6$ via T Duality,''
  Phys.\ Rev.\ Lett.\  {\bf 110}, no. 23, 231601 (2013)
  [arXiv:1212.1043 [hep-th]].


\bibitem{Sfetsos2014}
  K.~Sfetsos and D.~C.~Thompson,
  ``New ${\cal N} = 1$ supersymmetric $AdS_5$ backgrounds in Type IIA supergravity,''
  JHEP {\bf 1411}, 006 (2014)
  [arXiv:1408.6545 [hep-th]].

\bibitem{Gauntlett2004}
  J.~P.~Gauntlett, D.~Martelli, J.~Sparks and D.~Waldram,
  ``Supersymmetric AdS(5) solutions of M theory,''
  Class.\ Quant.\ Grav.\  {\bf 21}, 4335 (2004)
  [hep-th/0402153].
  
\bibitem{Lozano:2013oma} 
  Y.~Lozano, E.~O.~\'{O} Colg\'{a}in and D.~Rodr\'{i}guez-G\'{o}mez,
  ``{\em Hints of 5d Fixed Point Theories from Non-Abelian T-duality,}''
  JHEP {\bf 1405}, 009 (2014)
  [arXiv:1311.4842 [hep-th], arXiv:1311.4842].
  
\bibitem{Gauntlett2014}
  A.~Donos and J.~P.~Gauntlett,
  ``Flowing from AdS$_{5}$ to AdS$_{3}$ with T$^{1,1}$,''
  JHEP {\bf 1408}, 006 (2014)
  [arXiv:1404.7133 [hep-th]].

\bibitem{Hassan:1999bv} 
  S.~F.~Hassan,
  ``T duality, space-time spinors and RR fields in curved backgrounds,''
  Nucl.\ Phys.\ B {\bf 568}, 145 (2000)
  [hep-th/9907152].

\bibitem{Gauntlett2004a}
  J.~P.~Gauntlett, D.~Martelli, J.~Sparks and D.~Waldram,
  ``Sasaki-Einstein metrics on S**2 x S**3,''
  Adv.\ Theor.\ Math.\ Phys.\  {\bf 8}, 711 (2004)
  [hep-th/0403002].

\bibitem{Gauntlett2004b}
  J.~P.~Gauntlett, D.~Martelli, J.~F.~Sparks and D.~Waldram,
  ``A New infinite class of Sasaki-Einstein manifolds,''
  Adv.\ Theor.\ Math.\ Phys.\  {\bf 8}, 987 (2006)
  [hep-th/0403038].

\bibitem{Witten:1998xy} 
  E.~Witten,
  ``Baryons and branes in anti-de Sitter space,''
  JHEP {\bf 9807}, 006 (1998)
  [hep-th/9805112].


\bibitem{Gauntlett:2006ai} 
  J.~P.~Gauntlett, E.~O Colgain and O.~Varela,
  ``Properties of some conformal field theories with M-theory duals,''
  JHEP {\bf 0702}, 049 (2007)
  [hep-th/0611219].


\bibitem{Gevorgyan2013}
  E.~Gevorgyan and G.~Sarkissian,
  ``Defects, Non-abelian T-duality, and the Fourier-Mukai transform of the Ramond-Ramond fields,''
  JHEP {\bf 1403}, 035 (2014)
  [arXiv:1310.1264 [hep-th]].

\bibitem{Kelekci2014}
  \"{O}.~Kelekci, Y.~Lozano, N.~T.~Macpherson and E.~\'{O}.~Colg\'{a}in,
  ``Supersymmetry and non-Abelian T-duality in type II supergravity,''
  Class.\ Quant.\ Grav.\  {\bf 32}, no. 3, 035014 (2015)
  [arXiv:1409.7406 [hep-th]].


\bibitem{Macpherson:2013zba} 
  N.~T.~Macpherson,
  JHEP {\bf 1311}, 137 (2013)
  [arXiv:1310.1609 [hep-th]].








\bibitem{Maldacena1998}
  J.~M.~Maldacena,
  ``Wilson loops in large N field theories,''
  Phys.\ Rev.\ Lett.\  {\bf 80}, 4859 (1998)
  [hep-th/9803002].

\bibitem{Sonnenschein2000}
J. Sonnenschein,  ``\emph{What does the string/gauge correspondence teach us about Wilson loops?}'' (2000) 
[\emph{Lectures presented at Santiago de Compostela-$99$}] [hep-th/0003032].

\bibitem{Nunez2009}
  C.~Nunez, M.~Piai and A.~Rago,
  ``Wilson Loops in string duals of Walking and Flavored Systems,''
  Phys.\ Rev.\ D {\bf 81}, 086001 (2010)
  [arXiv:0909.0748 [hep-th]].


\bibitem{Klebanov2008}
  I.~R.~Klebanov, D.~Kutasov and A.~Murugan,
  ``Entanglement as a probe of confinement,''
  Nucl.\ Phys.\ B {\bf 796}, 274 (2008)
  [arXiv:0709.2140 [hep-th]].


\end{thebibliography}
\end{document}